\documentclass[conference]{IEEEtran}
\usepackage{subcaption}
\usepackage{amsmath}
\usepackage[xindy]{glossaries}
\usepackage{graphicx}
\usepackage{epstopdf}
\usepackage{amssymb}
\usepackage{amsthm}
\usepackage{cite}
\usepackage{enumerate} 
\usepackage{tikz}
\usepackage{fancyhdr}
\usepackage{lettrine}
\usepackage{svg}
\usepackage{bm}
\usetikzlibrary{patterns}
\usetikzlibrary{shapes,arrows}
\usepackage{verbatim}

\usetikzlibrary{positioning}
\usetikzlibrary{shapes.geometric}
\usetikzlibrary{shapes.misc}
\usepackage{booktabs}
\usepackage{algorithm}
\usepackage[noend]{algpseudocode}
 \usepackage{siunitx}

\usepackage[font=small,labelfont=]{caption}

\title{Estimating Timing Advance for Sub-THz Distributed Systems from Sub-10 GHz Channel State Information}
\author{Nishant Gupta$^{1,3}$, Muris Sarajlic$^2$ and Erik G. Larsson$^1$ \\
	${}^1$Department of Electrical Engineering (ISY), Link{\"o}ping University, Link{\"o}ping, Sweden \\
    ${}^2$Ericsson Research, Lund, Sweden \\${}^3$Department of Communications and Computer Engineering, LNMIIT Jaipur, India\\
	Email: nishantgupta.nic@gmail.com, muris.sarajlic@ericsson.com, erik.g.larsson@liu.se 
 \thanks{This work has received funding from the EU programme Horizon Europe (No. 101096302 – 6GTandem). Nishant Gupta is with the Department of Communications and Computer Engineering, LNMIIT Jaipur, India. He was with Department of Electrical Engineering (ISY), Link{\"o}ping University, Link{\"o}ping, Sweden, when this work was performed.}
 }

\begin{document}

\maketitle
\begin{abstract}
Dual-band wireless architectures transmit the control information over the sub-10 GHz while reserving sub-THz for high data rate links, offering notable capacity gains. However, a critical bottleneck in such systems is timing synchronization. Due to the narrow beams of the sub-THz radio units (RUs), when the dual-band user equipment (UE) rotates or moves, it becomes necessary to switch the transmission between the sub-THz RUs. This switching requires recalibrating the timing of uplink (UL) and downlink (DL) transmissions to prevent communication disruptions. Moreover, for sub-THz RUs, the method introduces significant overhead and latency, especially when switches are frequent. Leveraging the reliable sub-10 GHz band offers greater resilience to UE mobility, making it suitable for control signalling. Thus, in this paper, we propose a deep learning-based algorithm that infers the propagation delay from the sub-THz RUs to the UE using sub-10 GHz channel characteristics. The inferred delay is used for calculating the timing advance for UL transmissions without the need for two-way synchronization. Simulation results show that the RU switch can be made seamless at the physical layer, without incurring any synchronization-related latency.
\end{abstract}

\section{Introduction} 
Sub-terahertz (sub-THz) communications has been regarded as a promising technology for 6G, owing to its ability to provide extremely high data rate for particular applications, such as augmented reality (AR) or extended reality (XR) \cite{9269931}. However, sub-THz communication faces several challenges, including blockages and significant path loss \cite{9887921}. Additionally, the small wavelengths in the sub-THz band cause most objects, like human bodies or concrete walls, to create blockages and reflections. Therefore, it is necessary to have a dense and distributed deployment of sub-THz base stations, also known as radio units (RUs). Distributed sub-THz RUs implementation can be conveniently combined with a deployment operating at a lower frequency band, e.g., sub-10 GHz. Such a dual-band setup has several benefits, two of the most important being:
\begin{itemize}
\item sub-THz can be used for offloading high-throughput traffic from sub-10 GHz;
\item sub-10 GHz can provide side information and control information that can improve the stability of operation and reduce control signaling overhead at sub-THz.
\end{itemize}

An example scenario covering the second point above regards RU selection. The straightforward method for RU/beam selection is to conduct an exhaustive search on all the possible beam pairs between the sub-THz RUs and user equipments (UEs) \cite{7744807}. The beam selection process can be accelerated, thus reducing the signaling overhead in sub-THz, by optimizing the beam selection using the channel characteristics from sub-10 GHz \cite{ibbc,9050553,10292615,10577648}. For instance, the authors in \cite{ibbc} proposed a deep learning-based approach to infer a suitable sub-THz RU while incorporating the UE's orientation. 

Another more involved example is transmitting control information on the sub-10 GHz and allocating the sub-THz connection for the transmission of data exclusively. Due to the high directionality of the sub-THz, when the UE moves or rotates, it needs to switch from one sub-THz RU to another. Thus, UE needs to perform time synchronization with the new sub-THz  RU and determine the timing advance (TA). This is because the propagation delay from another sub-THz RU may be different from the previous serving sub-THz RU. Conventional synchronization involves transmitting the downlink (DL) synchronization signal and receiving the uplink (UL) synchronization signal to estimate the propagation delay. For sub-THz RUs, the method introduces significant overhead and latency, especially when switches are frequent. Moreover, the advanced use-cases like AR/XR for sub-THz will not tolerate high latency. These limitations become even more pronounced under UE mobility, where repeated synchronization signaling can significantly degrade system throughput.

Leveraging the reliable sub-10 GHz band becomes promising as it offers greater resilience to UE mobility, making it suitable for control signalling. The existing works on dual-band systems, such as \cite{9769897,10511063}, have largely focused on the beam steering and tracking. Moreover, the authors in \cite{9832505} proposed a sub-6 GHz synchronous mmWave communication scheduler, where the sub-6 GHz serves as a control channel for mmWave scheduling such that the network utility across mmWave is maximized. In \cite{10538322}, the authors designed a hybrid beamformer in the mmWave band using the sub-6 GHz channel information that can maximize the spectral efficiency. However, these approaches continue to rely on conventional UL/DL synchronization procedures at the high-frequency band, and therefore do not address the fundamental overhead imposed by synchronization signaling at sub-THz. 

In this paper, we depart from this conventional paradigm by eliminating the need for explicit UL and DL (two-way) signal transmissions for synchronization at sub-THz altogether. Instead, we exploit channel state information (CSI) available at sub-10 GHz to infer the propagation delay of the corresponding sub-THz link. By learning the non-linear relationship between low-band CSI and sub-THz timing offsets, the proposed deep learning–based framework enables accurate TA estimation prior to sub-THz data transmission. As a result,  the RU or UE can pre-advance its transmission timing.  Additionally, sub-10 GHz enables fast initial access at sub-THz, reducing overheads at sub-THz. To the best of our knowledge, achieving sub-THz timing synchronization by leveraging sub-10 GHz CSI has not been investigated in the current literature, thereby introducing a new dimension of cross-band cooperation beyond beam management and scheduling.
\begin{figure}
    \centering
    \includegraphics[width=1\linewidth]{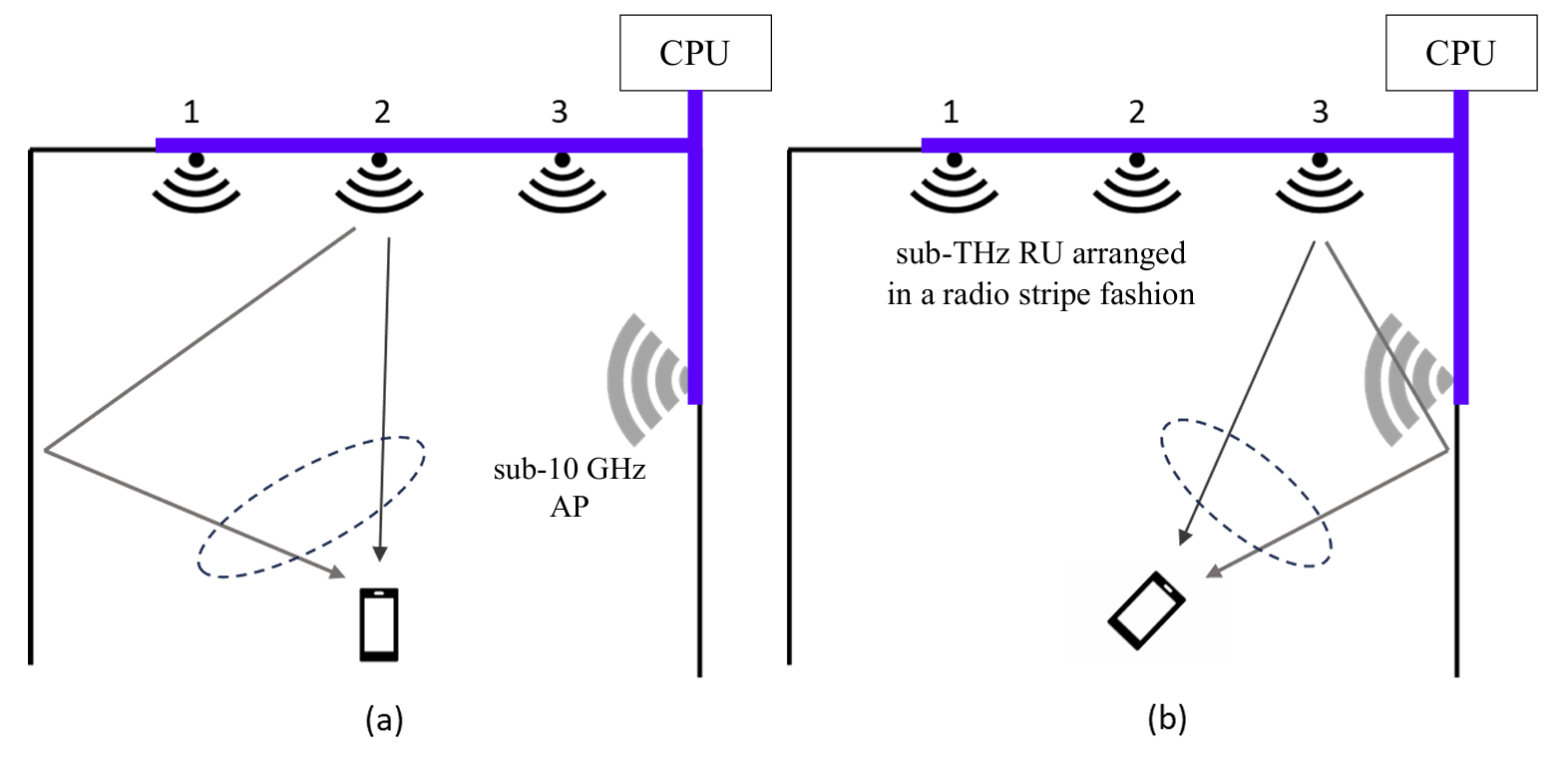}
    \caption{Illustration of the realignment between UE and sub-THz RU. }
    \label{fig:realignment}
\end{figure}
\section{System Model and Problem Formulation}
\subsection{System Model}
We consider an indoor communication system, where multiple sub-THz RUs (referred to as D-band) with multiple antenna elements are distributed and deployed alongside sub-10 GHz (called mid-band in 5G) APs with multiple antennas, as shown in Fig. 1. These APs and RUs are connected to the central processing unit (CPU). RUs are built into a dielectric fiber that can carry the RF propagation. All the sub-THz RUs are sequentially connected to each other, and the signal flow is unidirectional. Each sub-THz RU behaves as a distributed multiple-input multiple-output (MIMO) system, where the signal to each RU is generated in the CPU, which is then sent in analog form to each RU. 
\subsection{Problem Formulation}
\label{sec:Problem_formulation}
Due to high pathloss and low transmit power, sub-THz systems need to rely on beamforming in order to support a satisfactory link budget. This implies that the beams used, both at the network and UE side, will typically be narrow. As the UE moves or rotates, it will fall in and out of the coverage of individual RUs. This scenario is illustrated in Fig. \ref{fig:realignment}, where it is illustrated how the UE, due to rotation, falls out of the coverage of RU 2 and into the coverage area of RU 3. Such switches between RUs may happen often at sub-THz \cite{akyildiz2014terahertz}.
\begin{figure}[t]
    \centering
    \includegraphics[width=1\linewidth, trim={0 14cm 1cm 0},clip]{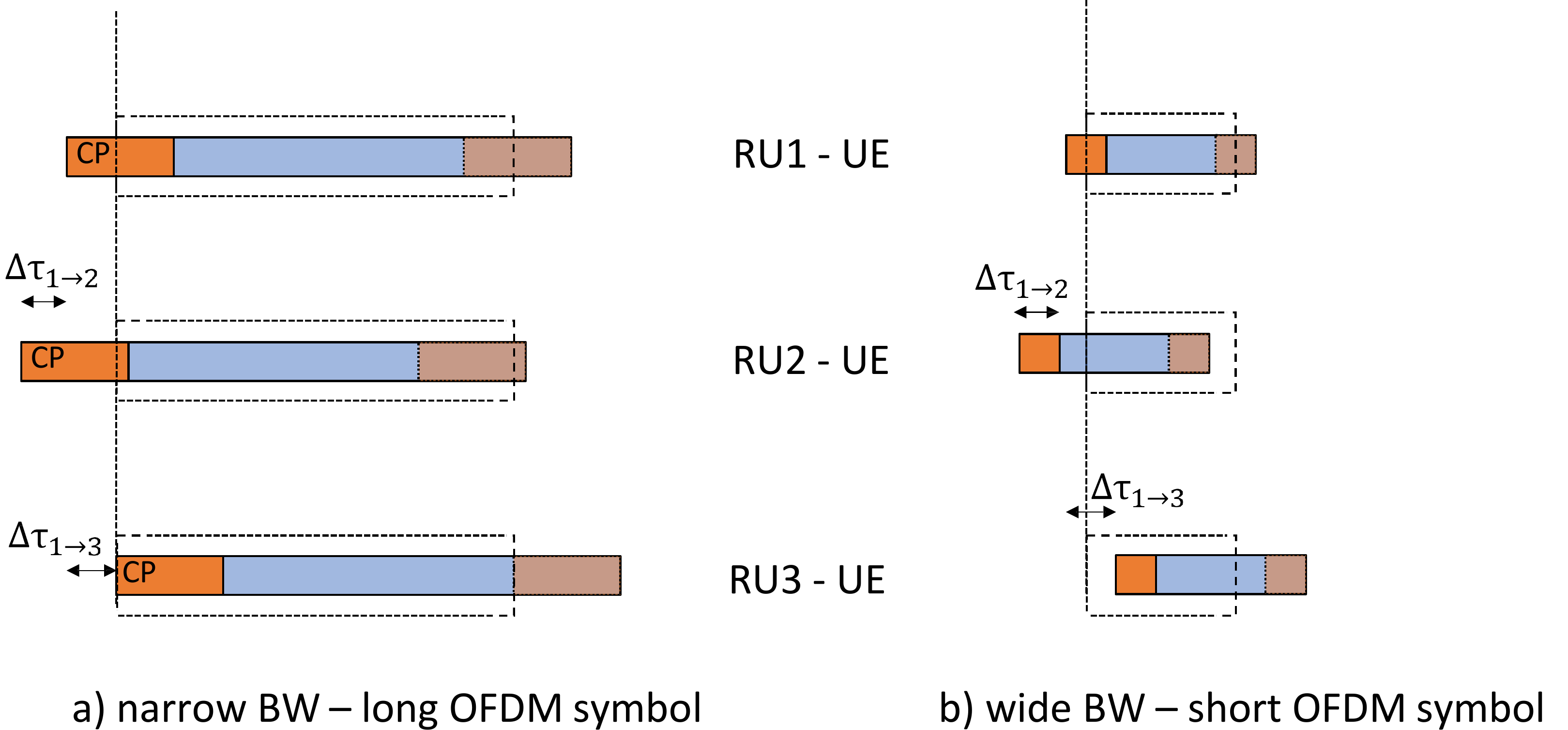}
    \caption{Time synchronization error and cyclic prefix. }
    \label{fig:FFT_window}
\end{figure}

\begin{figure*}[t]
    \centering
    \includegraphics[width=1\linewidth]{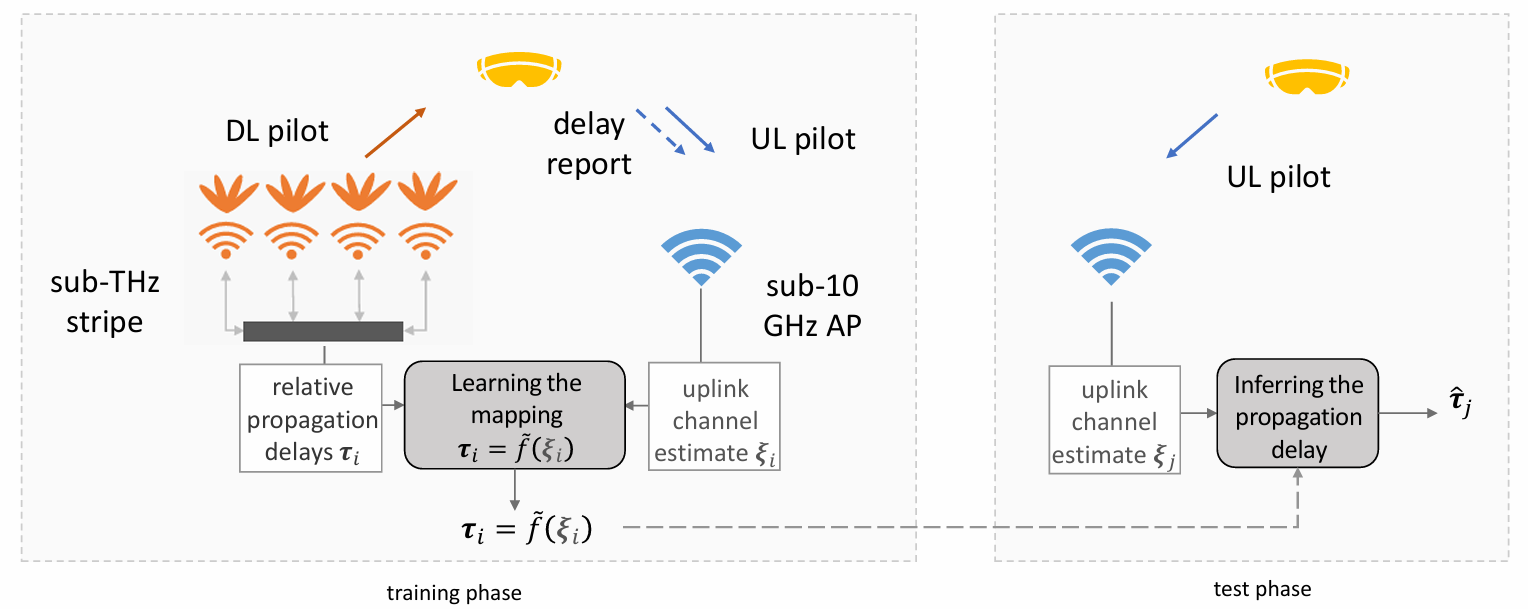}
    \caption{Illustration of training and testing phases in the proposed approach. }
    \label{fig:Training_and_testing_delays}
\end{figure*}
At lower frequencies, with narrower transmission bandwidths, the switch between serving RUs in a distributed system can be made seamless at the physical layer, without the need for resynchronization. Namely, when the physical layer is based on orthogonal frequency division multiplexing (OFDM) or related block-based transmission waveforms, any potential discrepancy in the RU-UE propagation delay (and hence time synchronization) between two RUs can be absorbed by the cyclic prefix (CP). The concept is illustrated in Fig. \ref{fig:FFT_window}(a), which illustrates the processes occurring at the UE receiver. It is assumed that the UE is time-synchronized with RU1 and that the start of the FFT window for the reception of the OFDM symbol (dashed box) is placed within the CP. In the time domain, the UE receiver sees a cyclic shift of the OFDM symbols, which translates to a phase ramp in the frequency domain - an effect easily removed by frequency domain equalization. Assume now that the serving UE switches to RU2 or RU3. Denote the propagation delays RU1 - UE, RU2 - UE and RU3 - UE by $\tau_{1}$, $\tau_{2}$ and $\tau_{3}$, respectively. Let the difference $\Delta \tau_{1 \rightarrow 2} = \tau_{1} - \tau_{2}$ be smaller than the duration of the CP as illustrated. If the UE keeps the placement of the FFT window according to the time synchronization with RU1, there will be a time synchronization error when switching to RU2. However, due to this error being smaller than the CP length, the UE receiver will simply see another cyclic shift of the OFDM symbol, the effects of which are again easily removed by equalization. Same holds for switching to RU3 from RU1, if $\Delta \tau_{1 \rightarrow 3} = \tau_{1} - \tau_{3}$ is smaller than the CP.

However, this synchronization error cannot be absorbed by the CP if the OFDM symbol and its CP are too short compared to the difference in propagation delays, as illustrated in Fig. \ref{fig:FFT_window}(b). In this case, a wrongly placed FFT window will incur a severe degradation in performance. Thus,  the UE must resynchronize with the new RU via DL reference signals for UE-side time synchronization, and the UE sends an UL reference signal for the purpose of network-side synchronization and determination of TA. The network then typically informs the UE of the TA so that future UL transmissions can be sent earlier, in order to align with symbol/slot/frame boundaries at the network side.

Operation at sub-THz assumes extremely large bandwidths, which implies very short CPs. Specifically, 6G TANDEM assumes operation in D-band with 7.5 GHz bandwidth. An OFDM waveform parameterization fitting this bandwidth allocation is an FFT size of 4k with a subcarrier spacing of 1920 kHz. Assuming a CP of length 288 (to keep CP overhead sathe me as in 5G NR), the length of the CP is 36.6 ns, which translates to a propagation distance of 10.9 meters. From the very simple calculation above, it follows that the inter-RU handover at sub-THz cannot be reliably made seamless at the physical layer, as described previously and illustrated in Fig. \ref{fig:FFT_window}(b), and explicit two-way resynchronization is required. Resynchronization when switching RUs would entail disrupting the DL transmission and increasing latency. It is therefore important to make the resynchronization process when switching between sub-THz RUs more efficient, shorter, or remove it altogether to ensure that latency is kept low and user quality of experience is high.  

\textit{Summary of problem:} Frequent sub-THz RU-UE switching necessitates resynchronization that results in excessive latency and control overhead. This raises a fundamental question: can the reliable sub-10 GHz band be exploited to provide TA for sub-THz links, thereby reducing or even eliminating the need for explicit two-way synchronization procedures at sub-THz?

\section{Proposed Methodology}
In this section, we first present the proposed approach, followed by dataset generation and deep neural network model.

\subsection{Deep Learning-based approach}
We propose a solution methodology that utilizes CSI obtained from sub-10 GHz to solve the resynchronization problem. Specifically, we propose a supervised learning approach wherein a mapping is learned between UE-to-network channel characteristics at sub-10 GHz and mean propagation delays from each of the sub-THz RUs to the UE. When trained, the proposed algorithm will take the sub-10 GHz channel characteristic as input and output estimates of mean propagation delays from the RUs to the UE. These estimates can in turn be used to 
\begin{itemize}
\item infer the TA without the need for DL/UL synchronization signal transmissions, and/or
\item pre-compensate the DL transmissions by the correct delay at the network side.
\end{itemize}

The proposed deep learning approach is illustrated in Fig. \ref{fig:Training_and_testing_delays}. In the training phase, we note the propagation delays ${\bm{\tau}}_i = [\tau_1,\ \tau_2\, ...\ \tau_{N}]_i^T$ from the UE to each of the $N$ RUs for position $i$. Mean delay $\tau_n$ is determined from the power delay profile (PDP) of the channel between the UE and RU $n$ as
\begin{equation}
\tau_n = \frac{\sum_{l=0}^{L_n-1}p_l^{(n)} \tau_l^{(n)}}{\sum_{l=0}^{L_n-1}p_l^{(n)}},
\label{eq:mean_delay}
\end{equation}
where $L_n$ is the number of discrete taps in the PDP, $p_l^{(n)}$ is the power and $\tau_l^{(n)}$ is the delay of the $l$-th tap.
\begin{figure}
    \centering
    \includegraphics[width=1\linewidth, trim={0 0.4cm 0 0},clip]{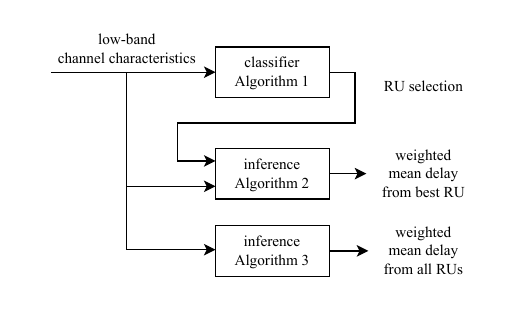}
    \caption{Proposed approach for the inference of propagation delay from sub-THz RU, where the algorithm is our proposed network. }
    \label{fig:algo_main}
\end{figure}
Simultaneously, the UL pilot is sent at sub-10 GHz, and the UL channel is estimated, from which a sub-10 GHz channel characteristic $\bm{\xi}_i$ is extracted. The channel characteristic $\bm{\xi}_i$ captures the temporal and angular characteristics of the channel, which contain information about the location of the UE. From a set of training data obtained by the UE or UEs occupying different positions $i$ in the coverage area, the model infers the mapping  ${\bm{\tau}}_i = \Tilde{f}(\bm{\xi}_i)$. In the test phase, the UE is located at a position $j$ and sends an UL pilot in sub-10 GHz, from which channel characteristic $\bm{\xi}_j$ is derived. Using the learned mapping $\Tilde{f}(\bm{\xi}_i)$, the model outputs estimated delays ${\bm{\hat{\tau}}}_j = [\hat{\tau}_1,\ \hat{\tau}_2\, ...\ \hat{\tau}_{N}]_j^T$ from the UE in position $j$ to all the sub-THz RUs. {\color{black} These delay estimates are subsequently used for TA compensation. For a given sub-THz RU $n$, the compensated TA is computed as
\begin{equation}
\text{TA}_{{n,j}^{\text{comp}}} = \hat{\tau}_n - \hat{\tau}_{\text{ref},j}, \label{eq:ta_comp}
\end{equation}
where $\hat{\tau}_{\text{ref},j}$ denotes a UE timing reference. The above equation applies the estimated one-way propagation delay as a relative timing offset, ensuring that the transmission arrives synchronously. Such delay-based TA compensation is consistent with established TA principles \cite{holma, 8643773}. Such computations are not feasible for real-time operation using conventional methods; therefore, a deep learning-based approach is employed to enable fast and accurate delay estimation.}

Fig. \ref{fig:algo_main} illustrates the proposed framework. Since the UE does not have prior knowledge of the sub-THz RU to which it will switch, we propose two variants of the proposed framework. In the first variant, the index of the best RU is estimated from channel characteristic $\bm{\xi}$ using a dedicated classification model referred to as Algorithm 1 in Fig. \ref{fig:algo_main}. The design and implementation of Algorithm 1 follow the methodology presented in \cite{ibbc}. Given the estimated best RU index and the sub-10 GHz channel characteristic, Algorithm 2 subsequently infers the weighted propagation delay to the selected \textit{best} sub-THz RU. In the second variant, referred to as Algorithm 3, the framework avoids RU selection and directly infers the weighted mean propagation delay across all candidate sub-THz RUs using the sub-10 GHz channel characteristics. 

\begin{figure}
    \centering
    \includegraphics[width=1\linewidth, trim={0 0.3cm 0 0},clip]{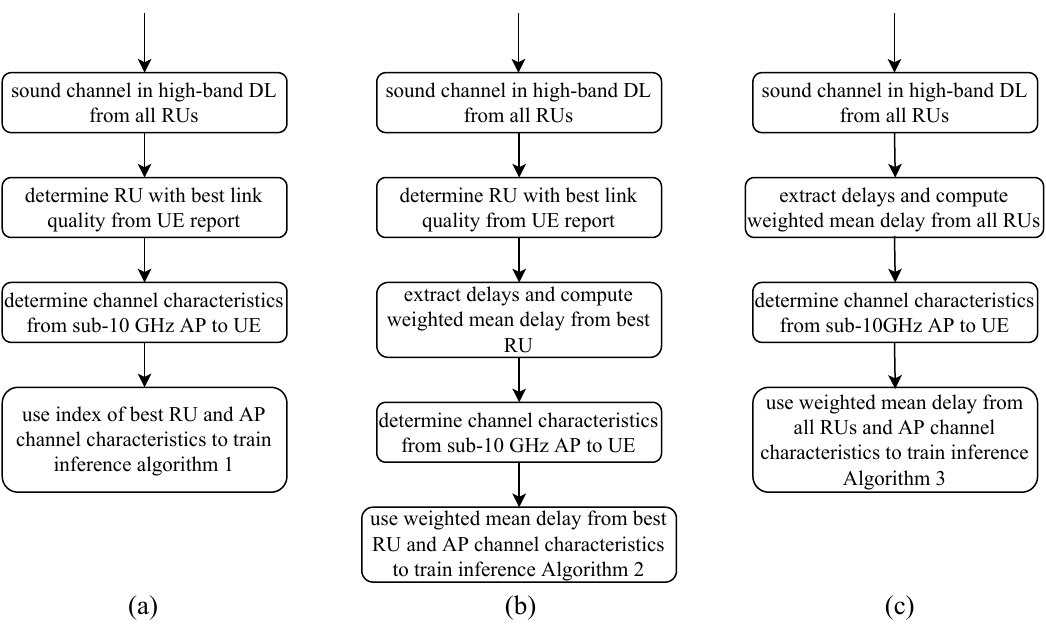}
    \caption{(a) Training of Algorithm 1, (b) Algorithm 2, and (c) Algorithm 3.}
    \label{fig:algo_ind}
\end{figure}

Fig. \ref{fig:algo_ind} illustrates the training procedures of three inference algorithms. Algorithm 1 (Fig. \ref{fig:algo_ind}(a)) uses high-band DL sounding to determine the best RU from the UE report. It uses the index of the best RU along with the sub-10 GHz channel characteristics to train the model. Algorithm 2 in Fig. \ref{fig:algo_ind}(b) computes the weighted mean delay of the best RU from (1) using high-band DL sounding, and similarly, Algorithm 3 (Fig. \ref{fig:algo_ind}(c)) computes weighted mean delay across all RUs; in both cases, the delays are combined with sub-10 GHz channel characteristics for training.

\subsection{Data Set Generation}
A ray tracing tool is used to construct the dataset using the shooting and bouncing rays method \cite{7152831}. We consider the office environment setup of dimension (8 x 5 x 5) meters (m) in length, width, and height, respectively, as described in \cite{MathWorks80211az2025}. It includes one sub-10 GHz and nine sub-THz RUs similar to \cite{tandem}, three stripes, each with three sub-THz RUs. The sub-10 GHz AP is placed on the eight-meter wall (in the middle), while RUs are positioned on the ceiling with a half-meter gap in between the RUs. The sub-10 GHz AP operates at 5.8 GHz with 100 MHz bandwidth, while the sub-THz RUs operate at 100 GHz with 7.5 GHz bandwidth, yielding delay resolutions of 10 ns and 0.13 ns, respectively. The height of the UE is restricted between 0.8 and 1.8 m. The algorithms are trained using the PDP. The PDP feature vector from sub-10 GHz has a size $N_f$ comprising path gains and path delays between the sub-10 GHz AP and UE. For Algorithm 1, the $N_f$ features serve as the input, and the corresponding labels represent the index of best RU $\in \{0, \cdots, 8\}$. For Algorithm 2, the same $N_f$ feature combined with the best RU index obtained from Algorithm 1, forms an ($N_f+1$) dimensional input, while the label corresponds to the true weighted mean propagation delay from the best RU computed via brute force. In Algorithm 3, only $N_f$ features are used as input, and the label corresponds to the true weighted mean propagation delay computed across all the sub-THz RUs ($N$ RUs) via brute force. 

Simulations are conducted in a single environment with a limited number of rays, enforcing a minimum inter-UE separation of 0.1 m, and users are assumed to be uniformly distributed. The simplified dataset is used to access feasibility and practicality of the proposed approach. Additional complexity, such as channel estimation and channel ageing effects, will be explored in future works. Training and testing datasets are generated separately within the same environment.
\begin{figure}[t]
\begin{subfigure}{\linewidth}
\includegraphics[width=1\linewidth]{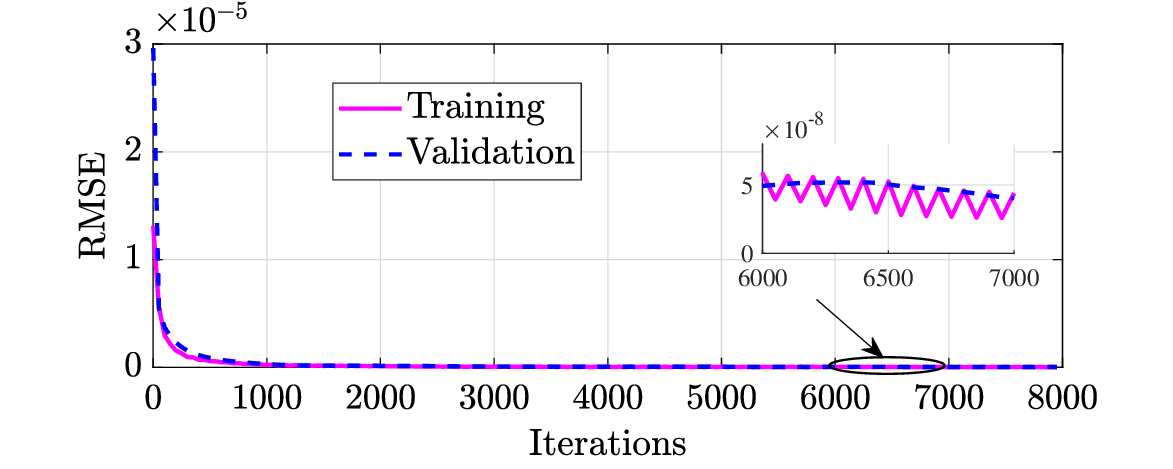}
\caption{Algorithm 2}
\end{subfigure}
\begin{subfigure}{\linewidth}
\includegraphics[width=1\linewidth]{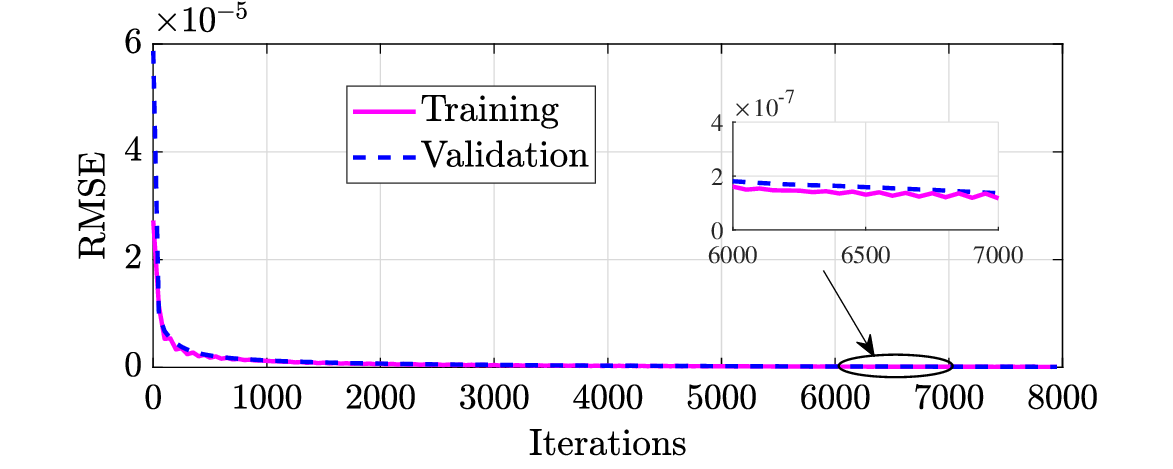}
\caption{Algorithm 3}
\end{subfigure}
\caption{Convergence performance of the loss function (RMSE).}
\label{fig:val1}
\end{figure}
\subsection{Proposed Deep Learning Model}
The proposed deep learning model (Algorithm 2 and Algorithm 3) is composed of an input layer, three hidden layers (with 64, 128, and 256 neurons), and an output layer. For Algorithm 2, the output has a single neuron acting as a single-output regression model. Furthermore, Algorithm 3 contains the multi-output of length $N$ behaving as a multi-output regression model. The learning rate is set to 0.01, and the Rectified Linear Unit (ReLU) activation function is used at all the hidden layers. Moreover, the activation function in the output layer is linear. The Stochastic Gradient Descent with Momentum (SGDM) optimizer is used, and the deep learning model is trained for 2000 epochs. \textit{Note:} In the proposed model, the total number of floating point operations (FLOPs) is 43352 FLOPs per iteration, and the approximate inference time for the network is $4.5 \times 10^{-4}$ seconds per input.

{\color{black} We train the deep learning model with appropriate training data until the loss is saturated.  Figs. \ref{fig:val1}(a) and \ref{fig:val1}(b) show the convergence performance of the loss function, i.e., root mean square error (RMSE), with the number of iterations (or 2000 epochs with 4 iterations per epoch). In particular, the RMSE of the training set and the validation set are compared in Fig. \ref{fig:val1}(a) for Algorithm 2 and Fig. \ref{fig:val1}(b) for Algorithm 3. It can be observed that the deep learning model converges quickly, and the difference between the training set and the validation set is quite small. }

\section{Results and Discussion}
In this section, we evaluate the performance of the proposed approach. Note that we have considered the inferred weighted mean propagation delay ($\hat{\bm\tau}$) as our output parameter as stated in Section III-A. 

\subsection{Performance Evaluation}
The performance of the proposed approach is shown in Fig. \ref{fig:diff}. The plot gives the CDF of inferred weighted mean delay for three different cases; (i) the difference of actual delay (obtained from brute force) from the best RU with that of the inferred delay from Algorithm 2, (ii) the difference of actual delay (obtained from brute force) from the best RU with that of the inferred delay from the Algorithm 3 (i.e., the best-inferred delay while considering all the RUs), and (iii) difference between the inferred delay from Algorithm 2 (inferred delay from best RU) and inferred delay from the Algorithm 3 (best while considering all the RUs). The distribution of the estimated propagation delay from different models is shown in Fig. \ref{fig:dist}. The results offer several interesting  insights:
\begin{itemize}
\item Estimating the delay to all the RUs has a lower prediction error (measured at the median and 90th percentile) than predicting the delays from the best RUs.
\item The prediction error of Algorithm 2 (measured at the median and 90th percentile) is below 10 ns. As a reminder, delay resolution at sub-10 GHz is 10 ns, and at sub-THz, it is 0.13 ns. Fusing the high delay resolution data from sub-THz and low delay resolution data from sub-10 GHz results in a delay prediction resolution that is higher than the lowest delay resolution in the dataset.
\end{itemize}

Most importantly, the results encourage an important practical application of the proposed method. Some steps of the explicit UL-DL synchronization between the UE and the new RU can be skipped. Let $\text{RU}_{old}$ denote the currently serving RU and $\text{RU}_{new}$ denote the RU that will take over as serving RU, and consider the possible applications: 
\begin{itemize}
\item In one application, the RU-UE delay for $\text{RU}_{new}$ inferred with the help of a sub-10 GHz UL pilot can be utilized to adjust the transmission time after the RU switch. Assume that the UE is time synchronized with $\text{RU}_{old}$ and that, through the inference process, the network estimates $\Delta \tau_{old \rightarrow new} = \tau_{old} - \tau_{new}$. The network may delay or advance the DL transmissions from $\text{RU}_{new}$ by $\Delta \tau_{old \rightarrow new}$ without informing the UE of the timing change. Likewise, the network may adjust its UL reception at $\text{RU}_{new}$ by $\Delta \tau_{old \rightarrow new}$. Neither DL nor UL synchronization signals need to be sent in this case, and the DL transmission does not have to be interrupted. {\color{black} If the estimated $\Delta \tau_{old \rightarrow new} << $ CP length, there will be no penalty in UL or DL performance.} Simulation results show that, for the considered scenario, the estimated $\Delta \tau_{old \rightarrow new}$ of 5 - 10 ns is much lower than the expected CP length of 36.6 ns as considered for the system calculation in Section \ref{sec:Problem_formulation}. In other words, the RU switch can be made seamless at the physical layer.
\begin{figure}
    \centering
    \includegraphics[width=1\linewidth]{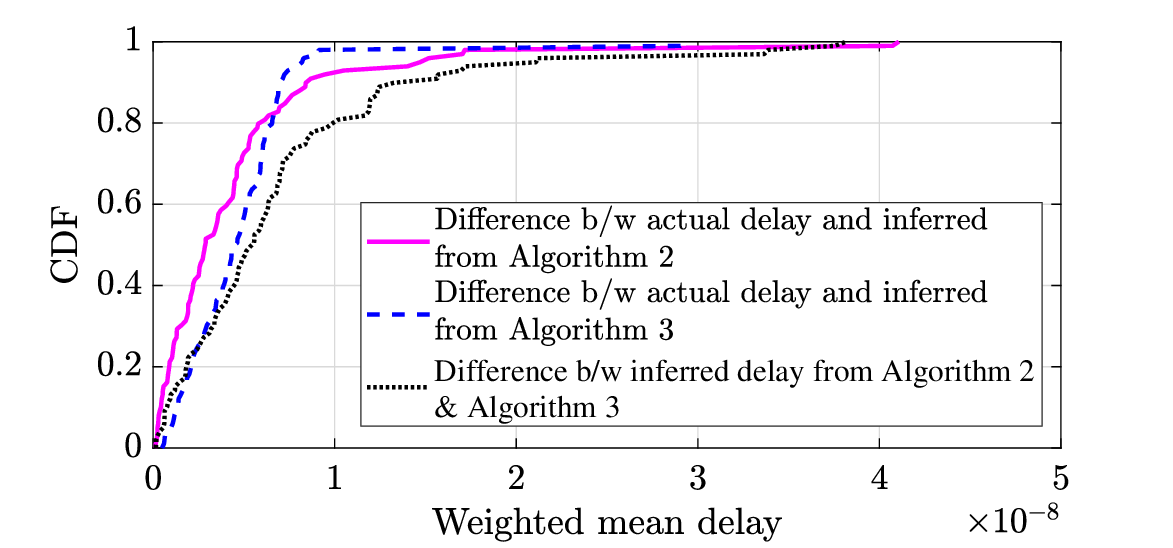}
    \caption{CDF of the inferred weighted mean delay.}
    \label{fig:diff}
\end{figure}
\item In another application, the network infers $\Delta \tau_{old \rightarrow new} = \tau_{old} - \tau_{new}$ and uses it to calculate a new TA value for communication with $\text{RU}_{new}$, which is signaled to the UE, preferably over the sub-10 GHz control link. The network will send a DL synchronization signal for the UE to readjust its time synchronization as the switch from $\text{RU}_{old}$ to $\text{RU}_{new}$ is performed. The TA value is used by the UE to advance its future UL transmissions. {\color{black} In this scenario, there is no need for the UL synchronization signal to be sent, and again, the DL transmission at sub-THz does not have to be interrupted.}
\end{itemize}
\begin{figure}
    \centering
    \includegraphics[width=1\linewidth]{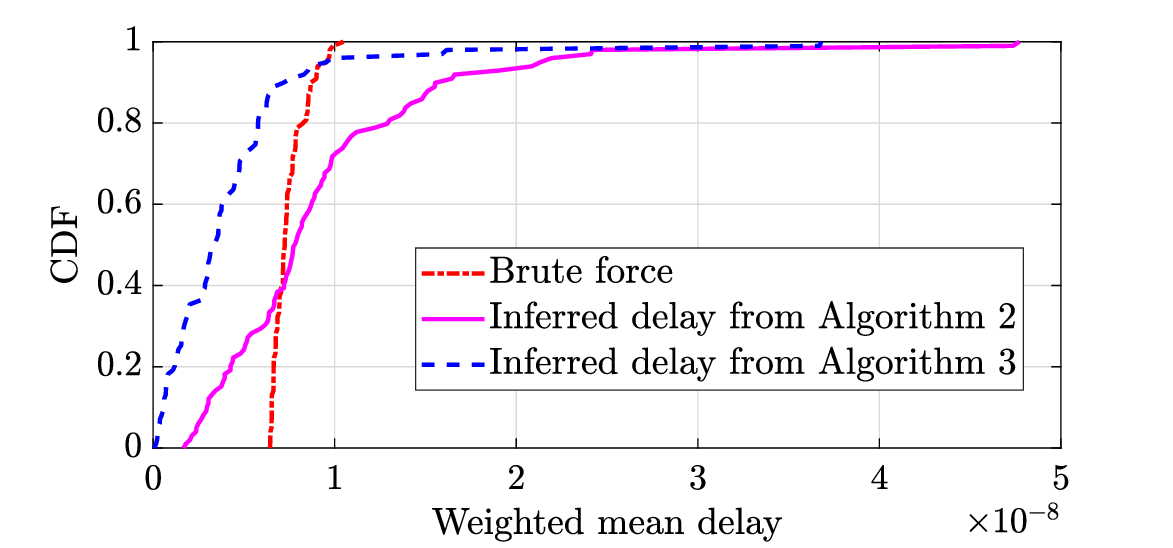}
    \caption{Distribution of weighted mean delay for different algorithms.}
    \label{fig:dist}
\end{figure}

\section{Conclusion}
Switching between sub-THz RUs to maintain communication as the UE moves/rotates poses a critical issue in the data transmission. This is because the propagation delay from another sub-THz RU may be different from the serving sub-THz RU. As a result, the signal may fall outside the CP, rendering the UE unable to decode it. Therefore, each switch requires recalibrating the timing to avoid potential disruptions. Traditional methods are resource-intensive. In this paper, we proposed a deep learning-based method to infer propagation delay from sub-THz RUs using sub-10 GHz channel characteristics. With our algorithm, a precise alignment of UE transmission timing across sub-THz RUs can be made reliably while reducing or eliminating the need for explicit two-way synchronization. Future work will extend the framework to more complex scenarios. 

\section*{Acknowlegement}
This work has received funding from the EU programmes Horizon Europe (No. 101096302 – 6GTandem). Nishant Gupta is now with the Department of Communication and Computer Engineering, LNMIIT Jaipur, India, was with the Department of Electrical Engineering (ISY), Link{\"o}ping University, Link{\"o}ping, Sweden, when this work was performed. 

\bibliographystyle{IEEEtran}
\bibliography{ref}

@online{tandem,
url = "https://horizon-6gtandem.eu/"
}

@ARTICLE{10538322,
  author={Huang, Zhaohui and Wang, Zhaocheng and Chen, Sheng},
  journal={IEEE Trans. Commun.}, 
  title={{Sub-6 GHz} Assisted {mmWave} Hybrid Beamforming With Heterogeneous Graph Neural Network}, 
  year={2024},
  volume={72},
month = {Nov.},
  number={11},
  pages={6917-6928},
  doi={10.1109/TCOMM.2024.3405372}}

@ARTICLE{9832505,
  author={He, Chenyuan and Wan, Yan and Zhao, Lu and Lu, Hongsheng and Shimizu, Takayuki},
  journal={IEEE Transactions on Vehicular Technology}, 
  title={{Sub-6 GHz V2X}-Assisted Synchronous Millimeter Wave Scheduler for Vehicle-to-Vehicle Communications}, 
  year={2022},
month = {Nov.},
  volume={71},
  number={11},
  pages={11717-11728},
  doi={10.1109/TVT.2022.3191423}}

@ARTICLE{8643773,
  author={Mukherjee, Sudarshan and Sinha, Alok Kumar and Mohammed, Saif Khan},
  journal={IEEE Trans. Commun.}, 
  title={{Timing Advance Estimation and Beamforming of Random Access Response in Crowded TDD Massive MIMO Systems}}, 
  month = {June},
  year={2019},
  volume={67},
  number={6},
  pages={4004-4019},
  doi={10.1109/TCOMM.2019.2900242}}

@book{holma,
  title={LTE for UMTS: OFDMA and SC-FDMA based radio access},
  author={Holma, Harri and Toskala, Antti},
  year={2009},
  publisher={John Wiley \& Sons}
}

@misc{MathWorks80211az2025,
  author       = {{MathWorks}},
  title        = {Three-Dimensional Indoor Positioning with {802.11az} Fingerprinting and Deep Learning},
  howpublished = {WLAN Toolbox Example, MathWorks Documentation},
  year         = {2025},
  note         = {Online. Available: https://in.mathworks.com/help/wlan/ug/three-dimensional-indoor-positioning-with-802-11az-fingerprinting-and-deep-learning.html},
}

@ARTICLE{7152831,
  author={Yun, Zhengqing and Iskander, Magdy F.},
  journal={IEEE Access}, 
  title={Ray Tracing for Radio Propagation Modeling: Principles and Applications}, 
month = {July},
  year={2015},
  volume={3},
  number={},
  pages={1089-1100},
  doi={10.1109/ACCESS.2015.2453991}}

@ARTICLE{9769897,
  author={Chafaa, Irched and Negrel, Romain and Belmega, E. Veronica and Debbah, Mérouane},
  journal={IEEE Trans. Wireless Commun.}, 
  title={{Self-Supervised Deep Learning for mmWave Beam Steering Exploiting Sub-6 {GHz} Channels}}, 
  year={2022},
month = {Oct.},
  volume={21},
  number={10},
  pages={8803-8816},
  doi={10.1109/TWC.2022.3170104}}

@ARTICLE{10511063,
  author={Zhao, Yao and Zhang, Xianchao and Gao, Xiaozheng and Yang, Kai and Xiong, Zehui and Han, Zhu},
  journal={IEEE Trans. Commun.}, 
  title={{LSTM-Based Predictive mmWave Beam Tracking via Sub-6 GHz Channels for V2I Communications}}, 
  year={2024},
month = {Oct.},
  volume={72},
  number={10},
  pages={6254-6270},
  doi={10.1109/TCOMM.2024.3395297}}

@INPROCEEDINGS{ibbc,
  author={Gupta, Nishant and Sarajlic, Muris and Larsson, Erik G.},
  booktitle={Proc. IEEE 101st Veh. Technol. Conf. (VTC2025-Spring), Oslo, Norway}, 
  title={{Deep Learning for sub-THz Radio Unit Selection Using sub-10 GHz Channel Information and Inferred Device Beamforming}}, 
month = {June},
  year={2025},
  volume={},
  number={},
  pages={1-5},
  doi={10.1109/VTC2025-Spring65109.2025.11174872}}

@article{akyildiz2014terahertz,
  title={Terahertz band: Next frontier for wireless communications},
  author={Akyildiz, Ian F and Jornet, Josep Miquel and Han, Chong},
  journal={Physical communication},
  volume={12},
  pages={16--32},
  year={2014},
month ={Sep.},
  publisher={Elsevier}
}

@ARTICLE{10292615,
  author={Vuckovic, Katarina and Mashhadi, Mahdi Boloursaz and Hejazi, Farzam and Rahnavard, Nazanin and Alkhateeb, Ahmed},
  journal={IEEE Trans. Wireless Commun.}, 
  title={{PARAMOUNT}: Toward Generalizable Deep Learning for {mmWave} Beam Selection Using Sub-6 {GHz} Channel Measurements}, 
  year={2024},
month ={May},
  volume={23},
  number={5},
  pages={5187-5202},
  doi={10.1109/TWC.2023.3324916}}

@INPROCEEDINGS{10577648,
  author={Pasic, Faruk and Hofer, Markus and Mussbah, Mariam and Caban, Sebastian and Schwarz, Stefan and Zemen, Thomas and Mecklenbräuker, Christoph F.},
  booktitle={Proc. Int. Conf. Smart Apps, Commun. Netw. (SmartNets), Harrisonburg, USA}, 
  title={Channel Estimation for {mmWave MIMO} Using Sub-6 {GHz} Out-of-Band Information}, 
  year={2024},
month ={July},
  volume={},
  number={},
  pages={1-6},
  doi={10.1109/SmartNets61466.2024.10577648}}

@ARTICLE{9050553,
  author={Uwaechia, Anthony Ngozichukwuka and Mahyuddin, Nor Muzlifah},
  journal={IEEE Access}, 
  title={A Comprehensive Survey on Millimeter Wave Communications for Fifth-Generation Wireless Networks: Feasibility and Challenges}, 
  year={2020},
month ={Mar.},
  volume={8},
  number={},
  pages={62367-62414},
  doi={10.1109/ACCESS.2020.2984204}}

@ARTICLE{7744807,
  author={Giordani, Marco and Mezzavilla, Marco and Zorzi, Michele},
  journal={IEEE Commun. Mag.}, 
  title={Initial Access in {5G mmWave} Cellular Networks}, 
  year={2016},
month ={Nov.},
  volume={54},
  number={11},
  pages={40-47},
  doi={10.1109/MCOM.2016.1600193CM}}

@ARTICLE{9887921,
  author={Shafie, Akram and Yang, Nan and Han, Chong and Jornet, Josep Miquel and Juntti, Markku and Kürner, Thomas},
  journal={IEEE Network}, 
  title={Terahertz Communications for {6G} and Beyond Wireless Networks: Challenges, Key Advancements, and Opportunities}, 
  year={2023},
month ={June},
  volume={37},
  number={3},
  pages={162-169},
  doi={10.1109/MNET.118.2200057}}

@ARTICLE{9269931,
  author={Petrov, Vitaly and Kurner, Thomas and Hosako, Iwao},
  journal={IEEE Commun. Mag.}, 
  title={{IEEE} 802.15.3d: First Standardization Efforts for Sub-{Terahertz} Band Communications toward {6G}}, 
  year={2020},
  month = {Nov.},
  volume={58},
  number={11},
  pages={28-33},
  doi={10.1109/MCOM.001.2000273}}

\end{document}